\documentclass[preprint2]{aastex}
\shorttitle{Two Energy Release Processes for CMEs}

\shortauthors{Chen, Hu {\&} Xia}

\begin{document}

\title{Two Energy Release Processes for CMEs: MHD Catastrophe and Magnetic Reconnection}

\author{Y. Chen\altaffilmark{1,2}, Y. Q. Hu\altaffilmark{2}, and L. D. Xia\altaffilmark{2}}

\altaffiltext{1}{Institute for Space Sciences, Shandong
University; Department of Space Science and Applied Physics,
Shandong University at Weihai, Weihai Shandong, 264209 China;
yaochen@sdu.edu.cn} \altaffiltext{2}{School of Earth and Space
Sciences, University of Science and Technology of China, Hefei
Anhui 230026, China}

\begin{abstract}
It remains an open question how magnetic energy is rapidly
released in the solar corona so as to create solar explosions such
as solar flares and coronal mass ejections (CMEs). Recent studies
have confirmed that a system consisting of a flux rope embedded in
a background field exhibits a catastrophic behavior, and the
energy threshold at the catastrophic point may exceed the
associated open field energy. The accumulated free energy in the
corona is abruptly released when the catastrophe takes place, and
it probably serves as the main means of energy release for CMEs at
least in the initial phase. Such a release proceeds via an ideal
MHD process in contrast with nonideal ones such as magnetic
reconnection. The catastrophe results in a sudden formation of
electric current sheets, which naturally provide proper sites for
fast magnetic reconnection. The reconnection may be identified
with a solar flare associated with the CME on one hand, and
produces a further acceleration of the CME on the other. On this
basis, several preliminary suggestions are made for future
observational investigations, especially with the proposed KuaFu
satellites, on the roles of the MHD catastrophe and magnetic
reconnection in the magnetic energy release associated with CMEs
and flares.
\end{abstract}



\section{Introduction}

Observations suggest that magnetic energy serves as a main energy
source for solar active phenomena such as CMEs (see reviews by
Forbes (2000) and Low (2001)), but it remains an open question how
the magnetic energy is released. The accumulated magnetic free
energy in the solar corona may be abruptly released either by a
global magnetic topological instability in a catastrophic manner
(e.g., Forbes and Isenberg, 1991; Isenberg et al., 1993; Forbes
and Priest, 1995; Hu et al., 2003) or by a fast magnetic
reconnection across preexisting or rapidly developing electric
current sheets (e.g., Antiochos et al., 1999; Forbes and Lin,
2000; Lin and Forbes, 2000). Although the two ways of energy
release are possible in the corona, the latter has often been
invoked in previous studies with implications in solar active
phenomena. However, such a mechanism is conditioned by the
preexistence or a rapid formation of electric current sheets. Then
a question remains: how can the current sheet exist stably or be
formed rapidly in the corona right before reconnection? Moreover,
a substantial heating of plasma must occur, but this is not a
general feature for CMEs (e.g., MacQueen and Fisher, 1983).
Therefore, it is necessary to find a mechanism that causes fast
release of magnetic energy without remarkable heating and leads to
a rapid formation and development of current sheets as well. The
first way of energy release mentioned above is exactly such a
mechanism.

Various theoretical models, including catastrophic models of
coronal flux ropes as well as other viable scenarios, were
proposed and used to simulate solar explosions such as flares,
prominence eruptions, and CMEs (see Forbes, 2000; Low, 2001; and
references therein). We give an overview, not intended to be
exhaustive, of the history and development of the catastrophe
models in this paragraph. For more details please refer to the
reviews written by Lin et al. (2003) and Hu (2005). To our
knowledge, the earliest flux rope catastrophe model is attributed
to Van Tend and Kuperus (1978) and Van Tend (1979) who
approximated the flux rope by a wire current filament and
concluded that a loss of equilibrium occurs if the current in the
filament exceeds a critical value. However, in their model and
subsequent similar ones, the field of the wire filament and the
background field are freely reconnected, so the ideal MHD
condition is disregarded. Soon their simple wire filament model
was refined and replaced by the so-called thin-rope model (e.g.
Forbes and Isenberg, 1991; Isenberg et al., 1993; Lin et al.,
1998), in which the ideal MHD condition is taken into account and
thus electric current sheets appear in the solution. The flux rope
is thin in the sense that its radius is far smaller than the
length scale of the photospheric field, an approximation purely
for analytical tractability. The thin-rope model was then extended
to numerical rope models (e.g., Hu et al., 2003), where the rope
is finite in radius. Such models were referred to as thick-rope
models by Hu (2005).

The studies of MHD catastrophe of coronal flux rope systems have
confirmed the possibility that the magnetic energy stored in the
corona is released by a global magnetic topological instability,
which is essentially an ideal MHD process. The instability takes
place in a catastrophic manner, and the plasma is accelerated by
the Lorentz force. As a result, the magnetic energy is mainly
transformed into the kinetic energy of plasma. In the meantime,
current sheets are bound to form as the eruptive flux rope drags
magnetic field lines outwards. In order for this mechanism to
work, one must find magnetic configurations with a catastrophic
behavior, and evaluate the magnetic energy of the system at the
catastrophe point, which is also called the energy threshold
representing the maximum magnetic energy that can be stored in the
system. There is no compelling reason that the threshold energy
for a catastrophe should exceed the open-field energy. If the
latter is larger, a CME-like expulsion is not expected unless
magnetic reconnection sets in to re-close part of the background
field that is opened up after a catastrophe (see, e.g., Lin and
Forbes, 2000). However, if the threshold energy is larger, then
there may be enough energy to open up the background field and
accelerate a CME out of the corona, simultaneously. That the
second possibility exists is significant as shown in, e.g., Hu et
al. (2003) and Li and Hu (2003).

We will summarize some recent results obtained in the study of
coronal flux rope catastrophe in the following with emphasis on
the catastrophic energy threshold. To further examine how the
magnetic energy be released during the catastrophe, a detailed
analysis of the force balance for the flux rope in either
equilibrium before or eruption after catastrophe is addressed in
section 3. To disentangle the contributions made by the ideal MHD
catastrophe and resistive magnetic reconnection to CME dynamics,
we construct a flux rope catastrophe model in the corona and solar
wind and compare different cases in which we either prohibit or
allow magnetic reconnection to take place across rapidly-growing
current sheets during the eruption. Finally, we conclude this
paper with several comments on how the future KuaFu mission (Tu et
al., 2007) may contribute to our understanding of the physics of
solar eruptive phenomena.

\section{Coronal flux rope catastrophe}

The so-called flux rope is defined as a twisted loop, a typical
structure in the corona. Theoretically, the flux rope must exist
for the support of prominences against gravity (Low and
Hundhausen, 1995), and it has two types of configurations, inverse
and normal, according to the types of associated prominences. To
our knowledge, most flux rope models so far belong to the inverse
type. It is presently impossible to directly observe the flux rope
in the corona. Nevertheless, Yan et al. (2001) claimed that they
found a flux-rope like structure in the corona through a
reconstruction of the coronal force-free field based on vector
magnetogram data observed at the photosphere.

For a magnetic configuration with an isolated flux rope, we may
introduce a set of parameters to characterize the properties of
the system. For the rope, one may take, say, its annular and axial
magnetic fluxes and the total mass in the rope. One may also
choose some appropriate parameters to characterize the background
field and plasma surrounding the flux rope of interest. All these
parameters are referred to as ``physical parameters''. On the
other hand, several parameters are introduced to describe the
geometrical features of the flux rope, for instance, the height of
the rope axis and the length of the vertical current sheet, which
is formed below the rope when it breaks away from the solar
surface and erupts upwards. Now we may select one of the physical
parameters as the control parameter that is changeable, and see
the variation of the geometrical parameters in response to the
change of the selected physical parameter. In studying the
parametric dependence of an equilibrium state, we often find that
a discontinuity can be encountered so that a small change in the
chosen parameter of variation produces an abrupt change in the
configuration of the equilibrium state. Then we say that the
system has a catastrophic behavior and identify the position of
the jump as the catastrophic point.

A catastrophe can be caused by a change of the rope properties, a
change of the background field or both. Let us give two typical
examples. The first is shown in Fig. 1, a thin-rope model in
Cartesian geometry developed by Forbes and Priest (1995), where
$\lambda$ is the half-distance between two point sources on the
photosphere, $h$ the height of the rope axis, and $R_0$ the radius
of the flux rope. The magnetic configurations shown in panels (1b)
- (1d), which are associated with different values of $\lambda$,
are plotted in the semi-infinite $x-y$ plane with the surface
$y=0$ corresponding to the photosphere. The catastrophe is caused
by a decrease of $\lambda$ to be caused presumably by converging
photospheric motions. The location where current sheets start to
form is pointed out in the left panel of this figure. It can also
be seen from this panel that the catastrophic curve of $h$ versus
$\lambda$ is S-shaped with a finite jump for $h$ from 1 to 9. The
second example is a thick-rope model, developed by Sun and Hu
(2005). The flux rope is embedded in a quasi-static helmet
streamer surrounded by a steady solar wind, and the catastrophe is
caused by a slight change of one of the physical parameters of the
flux rope. The flux rope stays in equilibrium before and erupts
upward after catastrophe. Fig. 2 shows an eruption of the flux
rope right after the catastrophic point in terms of the axial flux
of the rope. The rope breaks away from the solar surface and
erupts to infinity, forming a vertical current sheet below, as
mentioned above. The corresponding catastrophic curve, i.e., the
height of the rope axis versus the axial flux of the rope, turns
out to be fold-shaped in this case.

An important issue is the catastrophic energy threshold, which is
defined as the magnetic energy of the flux rope system at the
catastrophic point, as mentioned previously. Two decades ago, Aly
(1984) put forward a conjecture saying that the magnetic energy
that can be stored in a force-free field with given normal
component and at least one end of each field line anchored at the
solar surface can not exceed the open field energy with the same
normal component at the solar surface. The issue raised by this
conjecture is important to the catastrophe theory since one would
expect that the energy threshold is larger than the corresponding
open field energy so that after the background field is opened up
by the erupting flux rope, there is still a certain amount of
magnetic free energy left to produce a reasonable eruption, as
mentioned previously in the text. Nevertheless, the Aly conjecture
does not apply to the situations studied by most present flux rope
models which have been simplified as two-dimensional (2-D)
analyses. In these models, the field lines of the flux rope are
levitating in the corona and not anchored to the solar surface. We
point out in passing that an infinite amount of energy is required
to open up a closed magnetic field in 2-D Cartesian geometry (Hu
et al., 2003), therefore, it is energetically impossible to open
the overlying field and to let the flux rope escape to infinity
without magnetic reconnection, as demonstrated by previous
catastrophe models assuming Cartesian geometry (e.g., Lin and
Forbes, 2000). On the other hand, in the spherical geometry the
open-field energy is finite and it can be exceeded by the flux
rope system as already shown by many calculations (e.g., Weber and
Sturrock, 2001; Choe and Cheng, 2002; Hu et al., 2003; Li and Hu,
2003; Flyer et al., 2004; Sun and Hu, 2005; Zhang et al., 2005;
Peng and Hu, 2005; Ding and Hu, 2006; Chen et al., 2006a). Another
basic difference between 2-D Cartesian and spherical models from
the point of view of force analysis will be mentioned in Section
3. It was calculated that the energy threshold is larger than the
corresponding open field energy by about 8\% for coronal flux rope
systems either without (Li and Hu, 2003) or with a solar wind (Sun
and Hu, 2005) for a closed or partly open dipolar background
field. More careful analyses revealed that the energy threshold
depends slightly on the physical properties of the rope (Chen et
al., 2006a) and the background field (Peng and Hu, 2005; Ding and
Hu, 2006). Thus, in 2-D spherical geometry it is possible to have
the flux rope erupt to infinity when taking the ideal MHD
catastrophe as the only energy release process. This has been
confirmed with the numerical thick-rope models. It is also true
that the eruptive speed can be significantly enhanced after
magnetic reconnection sets in across the rapidly-developing
current sheets, as will be illustrated in the following section.

Based on the studies of MHD catastrophe of coronal flux rope
systems mentioned above, we argue that MHD catastrophe is probably
the main means of energy release for CMEs at least in the initial
phase. It releases energy without ohmic heating, especially
suitable for CMEs without associated flares. A by-product of the
catastrophe is the formation of one or more electric current
sheets, which proceeds at the Alfv\'enic time scale. This provides
proper sites for fast magnetic reconnection and answers the
question how current sheets are formed rapidly right before the
occurrence of magnetic reconnection. Such a reconnection further
releases the magnetic energy and should be responsible for a solar
flare associated with a CME event.

So far most flux rope models have been limited to 2-D analyses, as
mentioned previously. In 3-D cases, the two ends of a flux rope
are believed to be anchored to the solar surface. If the Aly
Conjecture is correct in this situation, the catastrophic energy
threshold must be less than the corresponding open field energy.
Magnetic reconnection is then necessary to make a catastrophe
develop into an eruption. So the catastrophe plays a role of
trigger for CMEs in this case. Nevertheless, Li and Hu (2003)
inferred that the Aly Conjecture may become invalid for systems
with catastrophic behavior. Such an inference deserves further
elaborations.

\section{Force balance of the rope in equilibrium or eruption and effects of reconnection on rope dynamics}

Now we turn to another important issue, the force balance problem
for the flux rope that is in equilibrium or eruption. Chen et al.
(2006b) made such an analysis for a flux rope embedded in either a
bipolar or a quadrupolar background field. Since the magnetic
energy is dominant over other forms of energy near the Sun, we
only analyze the interplay between different pieces of magnetic
forces, which are exerted by coronal currents inside and outside
the rope as well as the potential field with the same normal
component on the photosphere as the background field. For the
equilibrium situation, the resultant magnetic force acting on the
flux rope vanishes. On the other hand, if the rope erupts after
catastrophe, it was found that the resultant force is upward, and
thus the flux rope undergoes a continuous acceleration by the
Lorentz force. Fig. 3 shows the temporal profiles of various
magnetic forces acting on the flux rope and the resultant force
($\Sigma f$) as well during its eruption right after catastrophe.
The background field is a partly open bipolar field with an
equatorial current sheet extending to infinity, and magnetic
reconnection has been prohibited in both this sheet and the newly
formed current sheet below the erupting rope. These forces are
produced by the initial background potential field ($f_p$), the
azimuthal current in the rope and its image ($f_{R\varphi}$), the
poloidal current in the rope ($f_{Rp}$), the equatorial current
sheet above the rope inherent in the background field ($f_{c1}$),
and the newly formed vertical current below the rope ($f_{c2}$).
We emphasize that the self-interaction of the azimuthal current
inside the rope by itself results in an outward radial force on
the rope. This force comes from the curvature of the rope
surrounding the Sun, which is called the toroidal or "hoop'' force
by Chen (1989) and Krall et al. (2000) and the rope curvature
force by Lin et al. (1998). Note that in the 2-D Cartesian models
this self-force is trivially zero by the symmetry of an infinitely
long straight current, another basic difference between 2-D
Cartesian and spherical models as mentioned previously. As clearly
seen from Fig. 3a, the primary lifting force is $f_{R\varphi}$
whereas the primary pulling force is $f_p$. Fig. 3b is a local
enlargement of Fig. 3a to illustrate clearly the contributions of
$f_{Rp}$, $f_{c1}$ and $f_{c2}$ and the variation of the resultant
force. After about 20 minutes $\Sigma f$ changes from nearly zero
to positive, leading to a remarkable acceleration of the erupting
flux rope. Notice that the newly formed current sheet provides an
additional pulling force. Consequently, a weakening and
suppression of the current sheet by reconnection leads to a
further acceleration of the flux rope, as confirmed by the
following calculations.

As a first step to disentangle the contributions made by the ideal
MHD catastrophe and resistive magnetic reconnection to CME
dynamics, Chen et al. (2007) constructed a flux rope catastrophe
model in the corona and solar wind and compared different cases in
which magnetic reconnection is either prohibited or allowed to
take place across the rapidly-developing current sheets. For
simplicity, a polytropic process with the polytropic index
$\gamma=1.05$ is assumed to produce the background corona and
solar wind solution. One result of this model is presented in Fig.
4 for the case with the magnetic field strength at the polar hole
on the solar surface taken to be 16 G. The figure shows the
velocity profiles of different parts of the flux rope system,
including the cusp point (in dotted), the rope top (in dashed),
the rope axis (in solid), and the rope bottom (in dot-dashed),
thick curves for the reconnection case and thin for the case
without reconnection. It can be seen that the flux rope undergoes
an initial slow acceleration, followed by a fast one, and a slight
deceleration after it reaches a peak speed. The results are
essentially consistent with observed velocity profiles of CMEs
(e.g., Zhang and Dere, 2006). Comparing the solutions for the case
with and without magnetic reconnection, we can see that CMEs, even
fast ones, can be produced taking the ideal MHD catastrophe as the
only process of magnetic energy release. Nevertheless, the
eruptive speed is significantly enhanced after magnetic
reconnection sets in.

\section{Suggestions on the diagnosis of energy release processes of CMEs with KuaFu}

Now let us come to the final topic: What can observers do with the
future KuaFu mission to clarify the roles of MHD catastrophe and
magnetic reconnection in energy release processes involved in
CMEs?

The KuaFu mission is designed to observe the complete chain of
space weather events from the solar atmosphere to geospace with
three satellites, including KuaFu-A at the L1 libration point
observing solar Hard X-ray, EUV and white-light emissions, radio
waves, local plasma and magnetic field, and energetic particles,
and KuaFu- B1 and B2 in elliptical polar orbits elaborated to
continuously observe the northern polar auroral oval (Tu et al.,
2007). Although KuaFu is still at its early stage of development,
various payload plans have been proposed. Among them, we are
particularly interested in the following ones mounted at KuaFu-A:
an EUV Disk Imager (EDI), a Multi Order Solar EUV Spectrograph
(MOSES for KuaFu), a Hard X-Ray and Gamma-ray Spectrometer (HXGR),
a Lyman-$\alpha$ coronagraph, and a white light coronagraph. In
the following we briefly discuss some relevant characteristics of
these instruments and show how they may contribute to our
understanding of solar eruptions.

The EDI instrument will provide a continuous imaging in the
Lyman-$\alpha$ wavelength of 121.6 nm with high spatial and
temporal resolution. The polarization of this line will also be
recorded simultaneously. Complementary with ground-based or
space-borne magnetograms, these images with the deduced
polarization enable us to reveal the dynamical features and get
information about the associated magnetic topology in both large
and small scales before and after eruptions. With the
Lyman-$\alpha$ imaging channel of EDI (observing from the disk up
to 1.1 $R_\odot$), the Lyman-$\alpha$ (from 1.1 to 2.5 $R_\odot$)
and white-light coronagraphs (from 2.5 to 15 $R_\odot$), KuaFu can
provide a continuous tracking of a CME event from the disk source
to 15 $R_\odot$. The temporal profiles of the various parts of the
eruptive structure can be determined so as to put constraints on
the acceleration mechanism of individual CME events. These
observations are pressingly wanted by the solar physics community
especially after the failure of LASCO-C1 in the June of 1998. The
MOSES for KuaFu is a slitless imaging spectrograph at 3 spectral
orders in the He II 30.4 nm line providing high-resolution images
and simultaneous measure of the line of sight velocity on the
solar disk with an accuracy of 20 km s$^{-1}$. MOSES can be used
to find out the exact source region of the CME by e.g., detecting
outflowing materials in coronal dimming regions, and measure the
flux rope twist and CME velocity in the early phases of eruption.
It should be kept in mind that although this set of instruments is
designed to cover many aspects of a solar phenomena, the data set
they offer should be combined with complementary data from
instruments of other space crafts and ground based observatories.
For example, the K-Coronameter of the High Altitude Observatory in
Hawaii observes CMEs with a field of view from the limb to 2
$R_\odot$ in heliocentric distance, will play a complementary role
to the proposed KuaFu coronagraphs.

These new coordinated measurements by KuaFu are certainly
important to our understanding of the energy storage and release
processes, trigger, initiation and further acceleration of CMEs,
and will greatly facilitate our endeavor in evaluating the roles
of reconnection and catastrophe in CME energetics and dynamics. A
specific relevant observational task will be to evaluate the
variation rate of magnetic flux as the coronal dimming region
forms and disappears in a CME event, and to see how they are
related to the CME kinematics. The latter rate is supposed to
represent the total magnetic reconnection rate associated with the
formation of the post-flare loops and giant X-ray arches observed
in the lower corona (Forbes and Lin, 2000). The work along this
line has been carried out by several authors with SOHO
measurements, e.g., Jing et al. (2005) and Qiu and Yurchyshyn
(2005). With the MOSES for KuaFu to detect the coronal dimming or
the source region connected to the CME, the EDI to measure the
polarization of the Lyman-$\alpha$ line which contains information
on the coronal magnetic field vector, and the HXGR to assess the
timing of reconnection, it is hopeful to obtain a more accurate
description of the reconnection rate, which can be further
employed to constrain theoretical endeavors in evaluating the
roles of reconnection and catastrophe in CME energetics and
dynamics.

\acknowledgements
This work was supported by the National Basic
Research Program of China (2006 CB806304) and the National Natural
Science Foundation of China (40404013, 40574064, and 10233050).

\begin{figure}
\epsscale{.8} \plotone{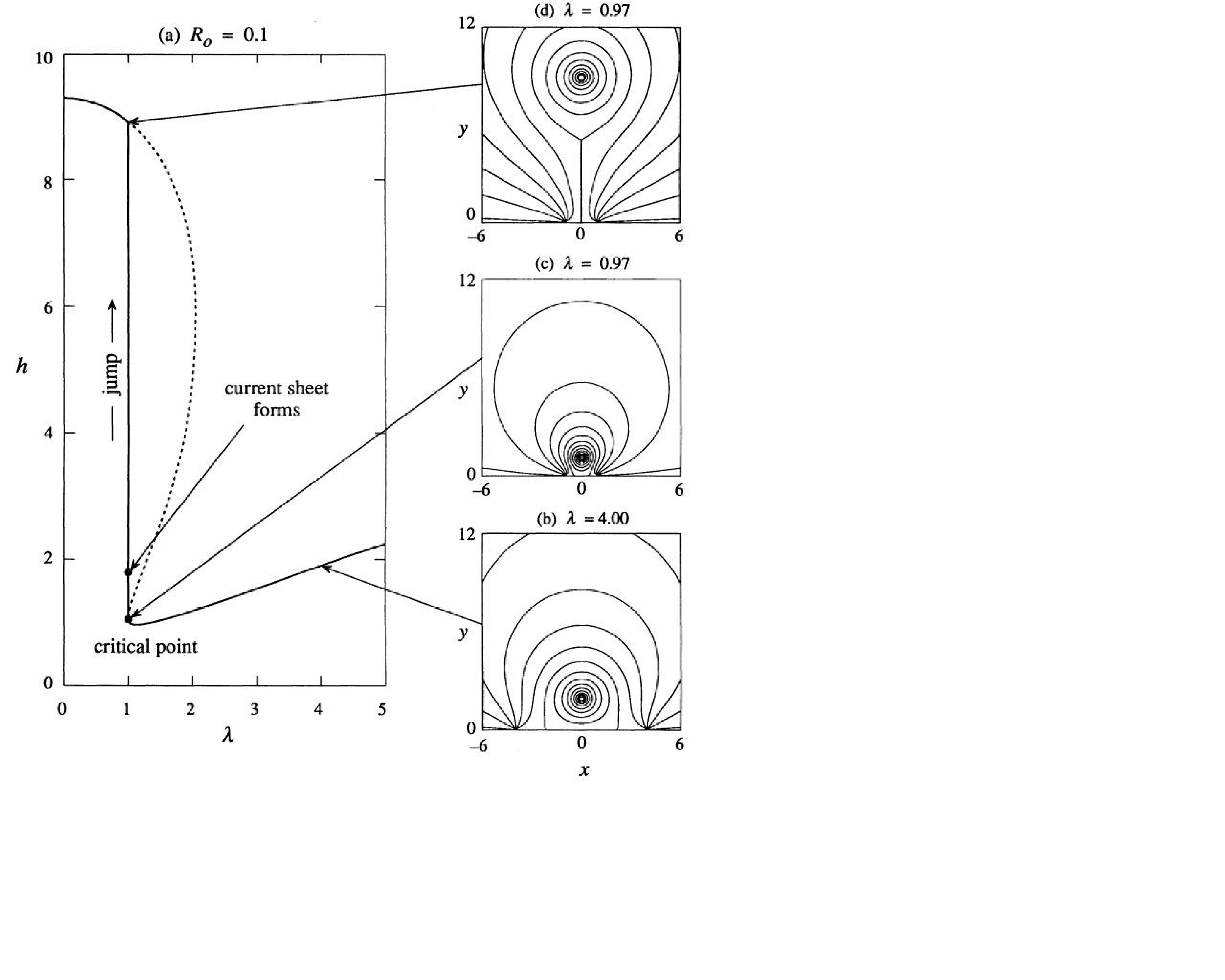} \caption{(a) Flux rope height, $h$,
as a function of the separation half-distance, $\lambda$, between
the photospheric sources. $R_0$ represents the radius of the flux
rope. Panels (b), (c) and (d) show magnetic configurations at the
3 locations indicated in (a) [after Forbes and Priest, 1995].}
\label{fig1}
\end{figure}

\begin{figure}
\epsscale{1.} \plotone{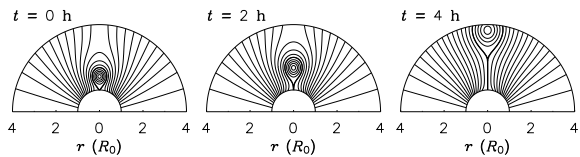} \caption{Magnetic configurations at
three separate times, showing an eruption of the flux rope right
after the catastrophic point [after Sun and Hu, 2005].}
\label{fig2}
\end{figure}

\begin{figure}
\epsscale{1.} \plotone{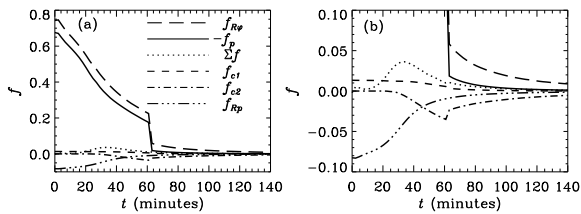} \caption{Temporal profiles of
magnetic forces on the eruptive flux rope per radian for the
bipolar background field case [after Chen et al., 2006b].}
\label{fig3}
\end{figure}

\begin{figure}
\epsscale{.6} \plotone{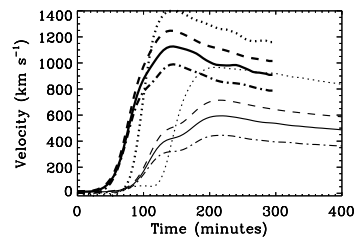} \caption{Temporal profiles of
velocity for the cusp, the rope top, the rope axis, and the rope
bottom in succession from higher to lower, thick curves for the
reconnection case and thin for the case without reconnection. }
\label{fig4}
\end{figure}

\end{document}